\documentclass[]{iopart} 
\begin{document} 

\newcommand{\beq}{\begin{equation}}
\newcommand{\eeq}{\end{equation}}
\newcommand{\beqn}{\begin{eqnarray}}
\newcommand{\eeqn}{\end{eqnarray}}
\newcommand{\dd}{\mbox{d}}
\newcommand{\dds}{\mbox{\scriptsize d}}
\newcommand{\dps}{\mbox{dPS}}
\newcommand{\smc}{\scriptsize}
\newcommand{\mycal}{\mathcal}
\newcommand{\porder}[1]{\mbox{${\mycal O}(#1)$}}

\noindent
\begin{minipage}[t]{10cm}
\footnotesize\it
\noindent
Talk given at the Durham workshop 
on HERA physics 1998\\ 
\end{minipage}
\hfill
\hspace*{\fill}
\begin{minipage}[t]{2cm}
\footnotesize
PSI--PR/98--29\\
December 1998
\end{minipage}
\vspace{1.0cm}

\begin{center} 
\title{
Jets and fragmentation
}
 
\author{Dirk Graudenz\footnote{{\it Electronic mail address:}
{\tt Dirk.Graudenz@psi.ch},\\ 
{\ }\,\,\,{\it WWW URL:} {\tt http://www.hep.psi.ch/graudenz/index.html}.
}} 
 
\address{Paul Scherrer Institut, 5232 Villigen PSI, Switzerland}
 
\begin{abstract} 
I review the status of next-to-leading-order calculations for hadronic
final states in deeply-inelastic lepton--nucleon scattering.
In more detail, I focus on 
calculations of (2+1)-jet-type cross sections, 
describe recent progress in extending the perturbative description
into the transition region between deeply inelastic scattering and 
photoproduction,
and study the validity of the fragmentation function picture
for one-particle-inclusive cross sections at small $Q^2$ and small~$x_p$.
\end{abstract} 
 
 

\end{center}
 


\section{Introduction} 
Recent experimental results from HERA at DESY show that the hadronic
final state in deeply inelastic scattering can be studied with high precision.
The results include the measurement of the strong coupling constant 
$\alpha_s(Q^2)$ by means of the (2+1)-jet rate 
$R_{2+1}=\sigma_{2+1}/\sigma_{\mbox{\smc tot}}$ \cite{1,2,3}
and event shapes \cite{4},
a direct determination of the gluon density \cite{5}, 
and the measurement of momentum fraction distributions for charged particles
\cite{6,7}. 
The latter indicates that it may be possible to study
scaling violations of fragmentation functions, the virtuality $Q^2$ of
the photon being the relevant scale for the fragmentation process.

\medskip

In this review I will
concentrate on three selected topics:

\begin{itemize}
\item {\bf NLO calculations for jet quantities:\\}
To exploit the increased experimental precision reliable theoretical
predictions in next-to-leading order (NLO) of QCD perturbation theory
are required. In this proceedings contribution I give an overview of recent
developments in NLO calculations for deeply-inelastic processes.
The main improvement during the last two to three years was that 
universal Monte Carlo programs have become available which permit the
numerical calculation of any (2+1)-jet-like infrared-safe observable in NLO.
\item {\bf Matching of DIS and photoproduction:\\}
For photoproduction ($Q^2 \approx 0$) 
and deeply inelastic scattering (DIS, $Q^2 \gg \Lambda_{\mbox{\smc QCD}}$) 
it is well 
known how to calculate cross sections systematically in perturbation theory.
Recently, a formalism \cite{8,9} has been developed which permits 
calculations
in the transition region $Q^2\sim \Lambda_{\mbox{\smc QCD}}$.
\item {\bf One-particle-inclusive processes:\\}
The measurements of momentum fraction distribution mentioned above show a very 
good agreement of experimental data and theoretical predictions in 
next-to-leading order for moderately large $Q^2$ and $x_p$. However, 
the theoretical 
prediction breaks down both for small $Q^2$ and small $x_p$.  
\end{itemize}

For lack of space, I had to leave out many, if not most, interesting topics. 
For more detailed
information, I would like to refer the interested reader to the
proceedings of the DIS~98 workshop in Brussels \cite{10}.


\section{NLO calculations for jet quantities} 
\label{jetc}
One of the basic problems of perturbative QCD calculations is that
experimentally hadrons are observed in the final state while theoretical 
calculations yield results for partons. Moreover, 
not all observables can be calculated in perturbation theory in a 
meaningful way. In principle, there are two possibilities:
\begin{itemize}
\item
{\it Infrared-safe observables}, which are constructed such that
all soft and collinear singularities cancel among real and virtual
corrections or can be absorbed into redefined parton densities.
The same observable is then evaluated both for parton final states (theory
prediction) and hadron final states (experimental data), possibly
after the experimental data have been corrected for systematic errors.
\item
Alternatively, additional non-perturbative objects can be introduced, 
for instance {\it fragmentation functions}, which allow for a study of
one-particle-inclusive processes. Fragmentation functions have to be measured
and parametrized experimentally, and may serve to hide final-state
collinear singularities
which do not cancel because of integrations over restricted phase space regions.
\end{itemize}
In this section I consider the first possibility; one-particle-inclusive
processes will be treated in Section~\ref{opi}.

In QCD perturbation theory, expectation values for parton observables
are calculated as a phase space integral of a product of 
a differential parton cross section $\sigma^{(n)}\left(p_1, \ldots, p_n\right)$
for $n$-parton final states 
and an observable ${\mycal O}^{(n)}\left(p_1, \ldots, p_n\right)$:
\begin{equation}
\label{thf}
\langle {\mycal O} \rangle =
\sum_n \int \dps^{(n)} \, \sigma^{(n)}\left(p_1, \ldots, p_n\right) \,
{\mycal O}^{(n)}\left(p_1, \ldots, p_n\right).
\end{equation}
In next-to-leading-order calculations, there are three contributions
to be included:
\beq
\langle {\mycal O} \rangle=\sigma_{\mbox{\smc Born}} \, {\mycal O}^{(n-1)}
+ \sigma_{\mbox{\smc virtual}} \, {\mycal O}^{(n-1)}
+ \sigma_{\mbox{\smc real}} \, {\mycal O}^{(n)}.
\eeq
Here $\sigma_{\mbox{\smc Born}}$ is the lowest order cross section, 
$\sigma_{\mbox{\smc virtual}}$ are the virtual and $\sigma_{\mbox{\smc real}}$
are the real corrections.
If the Born term has $n-1$ final-state partons, $\sigma_{\mbox{\smc virtual}}$
will also have $n-1$ and $\sigma_{\mbox{\smc real}}$ will have 
$n$ final-state partons. Infrared singularities arise in 
$\sigma_{\mbox{\smc virtual}}$ in the loop integrations and in 
$\sigma_{\mbox{\smc real}}$ in the phase space integration over
$\dps^{(n)}$. 

As already mentioned in the introduction, theoretical predictions 
for partons in the final state are infrared-finite only for a special
class of observables. The technical requirement for 
{\it infrared-safe observables} is that they behave well under soft and 
collinear limits:
\beqn
\!\!\!\!\!\!\!\!
{\mycal O}^{(n)}\left(p_1, \ldots, p_i, \ldots, p_n\right)
&{{} \atop {{\displaystyle\longrightarrow}
             \atop
            {\scriptstyle p_i \rightarrow 0}}}&
{\mycal O}^{(n-1)}\left(p_1, \ldots, \hat{p}_i, \ldots, p_n\right),
\\
\!\!\!\!\!\!\!\!
{\mycal O}^{(n)}\left(p_1, \ldots, p_i, \ldots, p_j, \ldots, p_n\right)
&{{} \atop {{\displaystyle\longrightarrow}
             \atop
            {\scriptstyle p_i \parallel p_j}}}&
{\mycal O}^{(n-1)}\left(p_1, \ldots, \hat{p}_i, \ldots,
\hat{p}_j, \ldots, p_n, p_i + p_j\right).\nonumber
\eeqn
Momenta denoted by $\hat{p}$
are to be omitted.

The main technical problem is the extraction of the infrared
singularities from the real corrections. It turns out that this can be done in 
an observable-independent way, such that it is possible to build Monte-Carlo
programs which are able to integrate arbitrary infrared-safe observables.
This can be done because the structure of QCD
cross sections in kinematical limits is known:
the factorization
theorems of QCD \cite{11} state that the structure of the
parton cross section $\sigma_{\mbox{\smc real}}$
for collinear and soft limits is of the form of a 
product of a singular kernel $K$ and the Born cross section $\sigma^{(n-1)}$:
\beq
\label{factorization}
\sigma^{(n)} {{} \atop {{\displaystyle\longrightarrow}
             \atop
            {\mbox{\smc soft/collinear}}}}
K\,\sigma^{(n-1)}.
\eeq
The product of $\sigma_{\mbox{\smc real}}$ and 
${\mycal O}^{(n)}$ thus behaves in a simple
way: the cross section goes over into a kernel
$K$ and the Born cross section, and the observable approaches the 
corresponding observable for Born term kinematics.
The kernel $K$ is independent of the phase space variables of the
$(n-1)$-particle phase space, and thus the phase space integration over the
corresponding variables can be performed analytically.

\subsection{Calculations}

Particularly interesting for phenomenological applications are processes
with 2+1 jets in the final state (which means that 2 jets are produced from
the hard scattering cross section, plus the remnant jet of the incident 
photon). 
First of all, in
leading order of QCD perturbation theory 
these processes are 
of ${\mycal O}\left(\alpha_s\right)$, 
and are thus suitable for a measurement of the strong coupling constant
$\alpha_s$. Moreover, the gluon density enters in leading order in
the so-called boson--gluon-fusion process. Therefore, this process can also be
used to measure the gluon density \cite{12,13}.
 
By now there are several calculations for (2+1)-jet processes
available with corresponding  
weighted Monte-Carlo programs:
\begin{itemize}
\item {\tt PROJET} \cite{14}: The jet definition is restricted to 
the modified JADE jet clustering scheme; 
the program is based on the calculation published
in Refs.~\cite{15,16,17}.
\item {\tt DISJET} \cite{18}: Again the jet definition is restricted to 
the modified JADE scheme; the program is based on the calculation 
in Refs.~\cite{19,20}.
\item {\tt MEPJET} \cite{21}: This is a program for the 
calculation of arbitrary
observables which uses the phase-space-slicing method. The 
corresponding calculation 
\cite{22} 
uses the Giele--Glover formalism \cite{23} 
for the analytical calculation of the
IR-singular integrals of the real corrections, and the crossing-function
technique \cite{24} 
to handle initial-state singularities. The latter requires
the calculation of ``crossing functions'' for each set of parton densities.
\item {\tt DISENT} \cite{25}: 
This program is based on the subtraction method. 
The subtraction term is defined by means of the dipole 
formalism\footnote{
The subtraction term is written as a sum over dipoles (an ``emitter'' formed
from two of the original partons and a ``spectator'' parton). Besides
the factorization theorems of perturbative QCD, the main ingredient is 
an exact factorization formula for the three-particle phase space, which allows
for a smooth mapping of an arbitrary 3-parton configuration onto the
various singular contributions.
}{}\,~\cite{26,27}. 
\item {\tt DISASTER++} \cite{28}: 
This is a {\tt C++} class library\footnote{
The acronym stands for ``Deeply Inelastic Scattering:
All Subtractions Through Evaluated Residues''.
Most of the program is written in {\tt C++}. A {\tt FORTRAN} interface
is available; thus there is no problem to interface the
class library to existing {\tt FORTRAN} code.
}~{}.
The subtraction method
is employed, and the construction of the subtraction term resembles
the method of Ref.~\cite{29}, i.e.\ it is obtained by the evaluation
of the residues of the cross section in the soft and collinear limits.
Double counting of soft and collinear singularities
is avoided by means of a general
partial fractions method.
\item {\tt JetViP} \cite{30}: 
This program implements the calculation of \cite{31}, which extends the
previous calculations into the photoproduction limit $Q^2\rightarrow 0$.
The calculation has been done by means of the phase space slicing method.
Up to now, the polarization of the virtual photon is restricted to be 
longitudinal or transverse.
\end{itemize}
The two basic approaches which are employed to extract the infrared
singularities from the real corrections are the {\it phase-space-slicing 
method} and the {\it subtraction method}. 
\begin{itemize}
\item The phase-space-slicing method splits up the full parton phase
space into two regions: a region $R$ where all partons can be resolved, 
and a region $U$ where two or more partons are unresolved. This split is
usually 
achieved by means of a technical cut parameter $s_{\mbox{\smc min}}$. 
Two partons
with momenta $p_1$ and $p_2$ are 
unresolved if their invariant mass $2p_1p_2$ is smaller than 
$s_{\mbox{\smc min}}$
and resolved if it is larger.
The integration over the resolved region $R$ can be performed safely 
by Monte Carlo integration, because all infrared singularities
are cut out by the phase space cut. The integration over the unresolved
region $U$ is divergent and cannot be performed numerically, but because of 
the constraint 
$2p_1p_2<s_{\mbox{\smc min}}$ the cross section factorizes 
(see Eq.~\ref{factorization}) in the limit $s_{\mbox{\smc min}}\rightarrow 0$.
This contribution is {\it approximated} by this limit. The integration over
the singular region can be done analytically, and the divergent parts can
be extracted. In the limit of $s_{\mbox{\smc min}}\rightarrow 0$ the
sum of the two integrals over $R$ and $U$ should approach
the integral over the full phase space. It should be kept in mind that
this convergence has to be checked explicitly by varying $s_{\mbox{\smc min}}$
and looking for a plateau in this variable.
\item A calculation using the subtraction method defines a subtraction
term $S$ which makes the integral 
$\int \dps \, \left(\sigma^{(n)}\,
{\mycal O}^{(n)} - S \right)$ finite. 
The original integral is, as an exact identity, rewritten as
\begin{equation}
\int \dps \, \sigma^{(n)}\,{\mycal O}^{(n)} 
= \int \dps \, \left(\sigma^{(n)}\,{\mycal O}^{(n)}-S\right)
+ \int \dps \, S.
\end{equation}
The first integral can be done by a Monte Carlo integration. For the term
$S$, the factorization from Eq.~\ref{factorization} holds exactly. As for
the phase-space-slicing method, the second term is integrated analytically.
No technical cut-off has to be introduced\footnote{This is, strictly speaking, 
not correct. A dimensionless cut $t_{\mbox{\smc cut}}$
of the order of $10^{-10}$ to $10^{-12}$ is used 
to avoid phase space regions where the subtraction no longer works because
of the finite precision of floating point numbers.}.
\end{itemize}
Both methods have their merits and their drawbacks. The phase-space-slicing
method is technically simple and can be easily implemented once the matrix 
elements for the real and virtual corrections are known. The main problem
is the 
residual dependence on the 
technical cut $s_{\mbox{\smc min}}$. The independence of numerical 
results from variations of this cut has to be checked; moreover, the
integration over the region $R$ mentioned above requires very high
statistics, because the integration region is close to the singular limit.
The subtraction method does not require a technical cut, but the construction
of the subtraction term $S$ is usually quite involved. If this can be afforded,
the subtraction method is the method of choice.

\subsection{Program comparisons}
It is interesting to compare the available universal Monte Carlo programs 
numerically to check whether all available calculations are consistent. 
Experimental papers usually contain statements that the programs 
``agree on the one per cent level''. A closer investigation, however, 
reveals that a statement of this kind is not correct.
The three programs {\tt MEPJET}, {\tt DISENT} and {\tt DISASTER++}
have been compared in Ref.~\cite{28} 
for the modified JADE jet clustering 
algorithm in the E-scheme for several choices of physical and unphysical
parton densities\footnote{By ``unphysical'' I mean parton densities
of the form $q(x)=(1-x)^\alpha$ and $g(x)=(1-x)^\alpha$, where
$\alpha$ is some power. These are introduced to have a more stringent test
on the hard scattering matrix elements.}. 
The result is that {\tt DISENT 0.1} and {\tt DISASTER++ 1.0}
agree well, with discrepancies of the {\tt MEPJET} results. Presently this
is studied in the framework of the HERA Monte Carlo 
workshop\footnote{{\tt http://home.cern.ch/}$\sim${\tt graudenz/heramc.html}}
at DESY. There does not yet exist a systematic comparison of the {\tt JetVip}
program with {\tt MEPJET}, {\tt DISENT} and {\tt DISASTER++}.

\section{Matching of DIS and photoproduction}
\label{mdp}
For large photon virtuality $Q^2$, the coupling of the exchanged virtual
photon in a lepton--nucleon scattering process is exclusively 
pointlike. Extending this kind of calculation down to $Q^2\approx 0$ 
leads to the problem that the photon propagator diverges. Instead, it is
possible to calculate the scattering process for the scattering of a quasi-real
photon and a nucleon, where the flux of quasi-real photons is described
by a Weizs\"{a}cker--Williams approximation.
For small $Q^2$, 
in addition to the cross section contribution from
the pointlike coupling, a resolved contribution has to be added, because
the quasi-real
photon may fluctuate into a hadronic state, which in turn interacts 
strongly with the incident nucleon. This process is modelled by means
of parton densities $f_{i/\gamma}$ of the virtual photon. 
The assumption of a photon structure is also required in order to treat
collinear singularities arising from the splitting of the real 
photon via its pointlike coupling into a collinear quark-antiquark pair.
This collinear singularity does not cancel against the virtual corrections, 
but is absorbed into the $f_{i/\gamma}$. Typically, the infrared
singularities are regularized by dimensional regularization; the singularities
then show up as poles in $\epsilon$, where the space-time dimension is
set to $d=4-2\epsilon$.

This type of calculation has recently been extended to the case of 
exchanged photons with moderate $Q^2$ in 
Refs.~\cite{8,9}. 
Here, because $Q^2$ is finite, strictly speaking there is no 
collinear singularity, and therefore no poles in $\epsilon$ related to the
photon splitting arise. However, the integral over the 
phase space of the quark-antiquark pair yields a logarithm in $Q^2$.
Because this logarithm may be large, and can therefore spoil perturbation 
theory, it has to be resummed. This is done by absorbing it into the
redefined parton densities of the photon. The corresponding renormalization 
group equation then takes care of the resummation.

Depending on the factorization scale for the virtual photon, the resolved
contribution can be surprisingly large even for fairly large $Q^2$, compared
with the ``standard'' DIS calculation for the pointlike coupling. 
This seems to be
in contradiction with the statement that the resolved contribution should
die out for increasing $Q^2$. There are two reasons for this: (a)
The choice of the factorization scale $\mu$ for the resolved photon dictates 
the size of the resolved contribution. The parton density of the 
virtual photon is exactly zero for $\mu=Q$. The factorization scale employed
in Refs.~\cite{8,9} is $\sqrt{Q^2+E_T^2}$ ($E_T$ is the
transverse energy of the produced jets), which makes sure that even at 
large $Q^2$ there is always a resolved contribution.
(b) In the full NLO calculation, there are
four different matrix elements which contribute: the direct process in 
LO and NLO (with the dangerous logarithm in $Q^2$ subtracted), and the 
resolved process in LO and NLO.
It is expected that the sum of the first three processes reproduce the
result for the standard calculations, and this is indeed the case: the 
logarithm that has been subtracted for the direct coupling is added up
again via the parton density of the photon in the LO resolved contribution.
The difference comes from the resolved contribution in NLO: the corresponding 
parton subprocess, which is of ${\mycal O}\left(\alpha_s^3\right)$,
is one order in $\alpha_s$ higher than the ``standard'' calculation.
Thus, this contribution could be considered as part of the NNLO correction
to the Born term for (2+1)-jet production. Differences between the two 
approaches
are therefore expected.

\section{One-particle-inclusive processes}
\label{opi}
The comparison of $x_p$-distributions\footnote{The variable $x_p$
is defined to be the fraction $2E/Q$, where $E$ is the energy of an observed 
particle in the current hemisphere of the Breit frame.} 
for charged particle production from 
experimental data 
\cite{6,7}
and the NLO program {\tt CYCLOPS} 
\cite{32,33} leads to severe discrepancies for small values of
$Q^2$ or small $x_p$. For large $Q^2$ and large $x_p$, data and
theory agree nicely. Where does this discrepancy come from?

The theoretical prediction is made in the fragmentation function picture:
the cross section for the inclusive production of charged particles
is obtained by a convolution of the hard scattering cross section
calculated in perturbative QCD and fragmentation functions which have
been obtained from fits to $e^+e^-$ data. Fragmentation functions depend
on the momentum fraction $z$ of the parent parton carried by the 
observed particle. An assumption in this picture is that the mass
of the observed particle can be neglected relative to any other scale 
of the process, in particular relative to its momentum. The variable $z$
can thus be defined either by means of fractions of energies or fractions
of momenta. In the real world, the observed particle {\it has} a mass, 
and this gives, thus, rise to an uncertainty in the theoretical description.
It is clear that mass effects will be important if $x_p={\mycal O}(2m_\pi/Q)$, 
$m_\pi$ being a typical hadronic mass. It turns out that excluding 
data points with a value of $x_p$ close to or smaller than 
this leads to a good agreement
between data and theory. A different argument in terms of rapidities 
of partons and observed particles has been given in Ref.~\cite{34}.
During the Durham workshop, a quantitative estimate of power corrections
$\sim 1/Q^2$
to the fixed-order NLO prediction has been made.
Y.~Dokshitser and B.~Webber proposed a factor
$1/(1+4\mu^2/(x_pQ)^2)$, depending on a mass parameter~$\mu$, to be multiplied
with the NLO cross section; 
this factor together with a fit of $\mu$ is able to describe the
experimental data fairly well 
(see the contribution to these proceedings by P.~Dixon, D.~Kant, and 
G.~Thompson).

\section{Summary} 
I have discussed three topics related to hadronic final states at HERA.
For the basic processes, theoretical predictions are available in 
next-to-leading-order accuracy. Independent calculations permit the comparison
of results, and a few problems with Monte Carlo programs have already been
fixed. What is still missing are calculations for $W$ and $Z$-exchange 
in the subtraction formalism for jet cross sections. This is likely to
become available in the near future. Moreover, a calculation for transverse 
momentum spectra of charged particles has not yet been done. 
The calculation for the transition region of DIS and photoproduction
fills a gap in the theoretical description of lepton--nucleon scattering.
However, it is not yet clear whether the parton densities for virtual photons
are process-independent beyond NLO, such that they can be measured in
one process and used for predictions in a different one. The necessity to 
introduce a power correction term for one-particle-inclusive distributions
already at fairly large values of $Q^2$ shows that the calculation of
fixed-order QCD corrections is not sufficient for a good description of
experimental data. Unfortunately, a power correction term introduces an 
additional mass parameter, which cannot be calculated from first principles.

\section{Acknowledgements}
I would like to thank the organizers of the 3rd UK Phenomenology Workshop on 
HERA Physics
for the invitation
to Durham.
St.~John's College, where the workshop took place, really is an amazing place.
Discussions with 
P.~Dixon, G.~Kramer, B.~P\"{o}tter, 
M.~Seymour and G.~Thompson are gratefully acknowledged.

\section{References}

\newcommand{\bibitema}[1]{\bibitem{#1}}

\end{document}